\documentclass[runningheads]{llncs}
\usepackage[T1]{fontenc}
\usepackage{amsmath}
\usepackage[colorlinks=false,bookmarks=false]{hyperref}
\usepackage{booktabs}
\usepackage{graphicx}

\newcommand{\code}[1]{{\small{\texttt{#1}}}}

\begin{document}
\title{Wildcat: Educational RISC-V Microprocessors}
%
%
\author{Martin Schoeberl\inst{1}\orcidID{0000-0003-2366-382X}}
\authorrunning{Martin Schoeberl}
%
\institute{Department of Applied Mathematics and Computer Science\\
Technical University of Denmark, Kgs. Lyngby, Denmark\\
\email{masca@dtu.dk}}

\maketitle              
\begin{abstract}
In computer architecture courses, we usually teach RISC processors using a five-stage pipeline, neglecting alternative organizations. This design choice, rooted in the 1980s technology, may not be optimal today, and it is certainly not the easiest pipeline for education. This paper examines more straightforward pipeline organizations for RISC processors that are suitable for educational purposes and for implementing embedded processors in FPGAs and ASICs. We analyze resource costs and maximum clock frequency of various designs implemented in an FPGA, using clock frequency as a performance proxy. Additionally, we validate these results with ASIC designs synthesized using the open-source SkyWater130 process.

Contradictory to common wisdom, a longer pipeline (up to 5 stages) does not necessarily always increase the maximum clock frequency. In two FPGA and one ASIC implementation, we discovered that a four- or five-stage pipeline leads to a slower clock frequency than a three-stage implementation. The reason is that the width of the forwarding multiplexer in the execution stage increases with longer pipelines, which is on the critical path. We also argue that a 3-stage pipeline organization is more adequate for teaching a pipeline organization of a microprocessor.

\keywords{RISC processor architecture \and computer architecture education \and RISC-V.}
\end{abstract}

\section{Introduction}
\label{sec:intro}

When teaching computer architecture, we are used to teaching RISC processors' mechanics in a five-stage pipeline. We ignore entirely other organizations (except
teaching out-of-order execution in advanced computer architecture courses).

However, the organization of a RISC processor as a 5-stage pipeline is probably not the best today.
It stems from the early RISC processors in the 80' of the last century.
Today's technology probably favors different organizations.

In this paper, we explore different and simpler pipeline organizations for RISC processors
practicable for teaching computer architecture and probably also useful for the implementation
of embedded processors in field-programmable gate arrays (FPGAs) and in ASICs.

The RISC-V instruction set architecture (ISA) was defined by Andrew Waterman in
his PhD thesis~\cite{Waterman:EECS-2016-1} together with Dave Patterson and Krste Asanovic
at the University of California, Berkeley (UCB).
Andrew explored three decades of RISC architectures, e.g., MIPS, SPARC, Alpha, and others,
to extract the essence of an RISC instruction set. The strongest move in this project
was to provide an open-source ISA.
Open-source software has been the driver of the Internet development.
However, on the hardware side of computers, a processor's instruction set
is a very protected asset. The move to a free ISA will change the way
computers are built.

The RISC-V Foundation now maintains the RISC-V instruction set,\footnote{\url{https://riscv.org/}}
which has been established in 2015. The Foundation has more than
3,100 members across 70 countries. Besides managing the RISC-V ISA specifications,
the Foundation organizers regularly RISC-V workshops and summits.

This paper presents Wildcat, an open-source, educational implementation
of a RISC-V microprocessor. The name Wildcat was inspired by the nice running area
in Tilden Park in Berkeley: the Wildcat area. I started this project in 2015 with
a RISC-V ISA simulator after participating in Andrew's PhD defense and a run on the Wildcat Creek Trail.
An early version of Wildcat has been presented in~\cite{wildcat:2024}.

RISC-V is an ISA definition; it does not define an implementation.
The ``V'' stands for the fifth RISC project at UCB and also indicates
that vector instructions are a part of the standard.
However, in many textbooks, implementations of RISC-V
are presented as a 5-stage pipeline.\footnote{The author even witnessed a conference participant
arguing that the ``V'' in RISC-V mandates a 5-stage implementation.}

Educators, the author included, have taught computer architecture with a 5-stage pipeline organization only.
I regret my approach to teaching pipeline organizations has been too narrow.
This paper aims to argue for a simplified approach with fewer stages,
emphasizing efficiency and ease of understanding in contrast to the complexity of longer pipelines.

We start with a simple 3-stage pipeline and extend it to 4 and 5 stages.
We then compare these three designs developed in the same language,
in the same style, and programmed by the same author.
We will show that the additional resource usage of the additional stages does
not justify the marginal or non-existing performance gains.

We explore the resource costs and the maximum clock frequency
of the different organizations implemented in an FPGA.
As the clocks per instructions (CPI) are almost identical for the three designs,
the maximum clock frequency is a valid proxy for the performance of the
pipeline alone.
To validate the results in ASIC designs, we also synthesized the different versions with
the open-source OpenLane2 tools using the SkyWater130 process.
Our main finding is that a 3-stage pipeline organization is superior to a 4- or 5-stage organization.

The contributions of the paper are: (1) four clean implementations of RISC-V useful
for teaching computer architecture, (2) questioning the common wisdom to default
to a 5-stage pipeline, and (3) comparing 3, 4, and 5-stage pipeline implementations
written by the same person, in the same language, on the same FPGA. To the best of our knowledge, those
three organizations have never been compared directly.

This paper is organized into seven sections:
The following section presents the proposed pipeline organizations.
Section~\ref{sec:impl} describes our implementation of different pipeline organizations.
Section~\ref{sec:eval} evaluates the different organizations on two families of FPGAs and an ASIC synthesis flow.
Section~\ref{sec:related} presents related work.
Section~\ref{sec:conclusion} concludes.

\begin{figure*}[t]
  \centering
  \includegraphics[width=\textwidth]{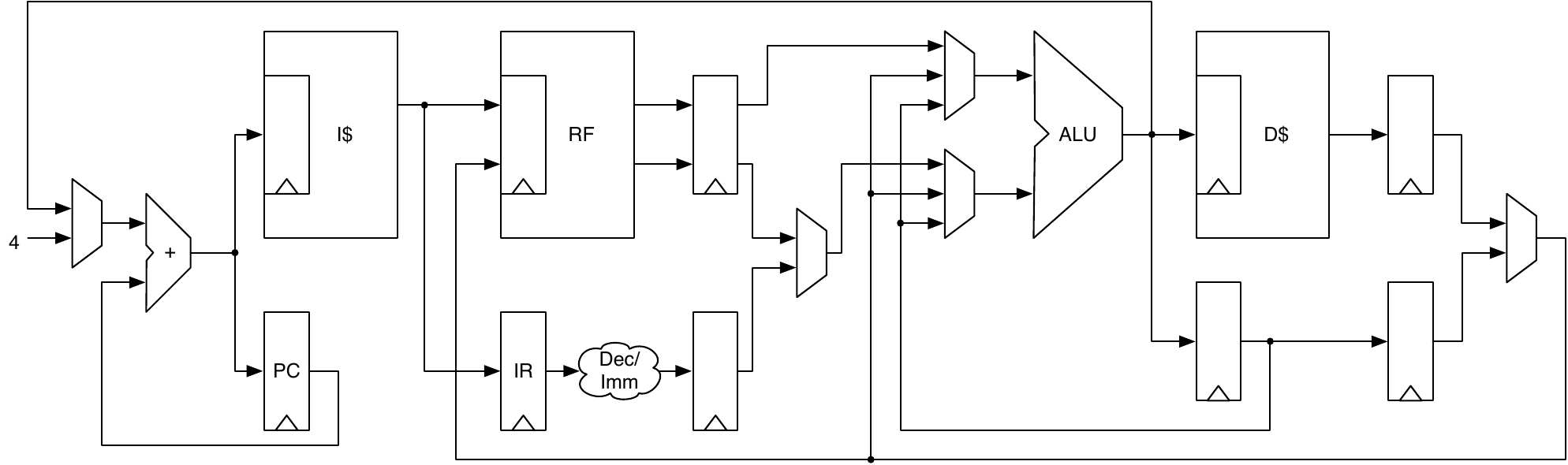}
  \caption{A textbook style 5-stages RISC-V processor pipeline.}
  \label{fig:5-stages}
\end{figure*}

\section{Pipeline Organizations}

In this section, we discuss three different organizations of in-order RISC pipelines.
We start with the classic, textbook style 5-stage organization and will argue step-by-step
to reduce it to three stages.
The arguments are based on my background in FPGA implementations.
However, in the evaluation section, we will also explore the three different organizations
in the open-source SkyWater130 ASIC technology.

\subsection{The Classic 5-Stage Pipeline}

The standard textbook organization~\cite{Hennessy06} \cite{harris2021digital} of a RISC pipeline is a
five-stage pipeline with the following stages: (1) instruction fetch (IF), (2) instruction decode and register
file read (ID), (3) execute (EX), (4) memory access (MEM), and (5) write back (WB).
Between each pair of stages, there are the so-called pipeline registers.
Figure~\ref{fig:5-stages} shows the simplified schematic of the classic 5-stages RISC pipeline.
The figure also shows the forwarding paths from MEM and WB to the input of the ALU in the EX stage.

The instruction cache (I\$), the register file (RF), and the data cache (D\$) can be implemented
with synchronous on-chip memories. Synchronous memories have registers at their address
and date inputs, and a read operation takes one clock cycle. Therefore, their input registers
are part of the pipeline registers between the different stages.

We are all so used to this textbook organization that we seldom question the number
of stages. Even Wikipedia does not discuss any other possible organization on the
page describing a RISC pipeline.\footnote{\url{https://en.wikipedia.org/wiki/Classic_RISC_pipeline}}
This paper explores different organizations, especially when
implementing that RISC pipeline in an FPGA.

In FPGAs, synchronous memories are fast.
They are cost-efficient compared to flip-flops. Therefore, the register file should be implemented using synchronous memory. Multiplexers, used for forwarding, are slow compared to on-chip memory in an FPGA.
From an earlier design of the Patmos~\cite{patmos:rts2018}
RISC processor, we know that the forwarding path in front of the
arithmetic logic unit (ALU) becomes the critical path.

\subsection{Do We Need a Write Back Stage?}

For a RISC implementation in current technology, we assume for the register file
either an implementation in flip-flops or using a synchronous on-chip RAM.\footnote{We
are aware of other organizations for a register file, e.g., with latches. However,
latches are impractical for FPGAs. We might explore other organizations in the future
with the SkyWater130 technology. Initial experiments to build a latch based register file
failed with OpenLane as yosys did not like the Verilog code for latches.} Therefore, the input to the register file
is the input of a flip-flop.

The write-back stage contains only a 2:1 multiplexer between the forwarded result
from the EXE stage and the result of a memory read.
The multiplexer sits between the MEM/WB pipeline register and the
input of the register file.

This stage performs very little work. However, that 2:1 multiplexer needs
to be forwarded to the EXE stage, adding yet another multiplexer to the
ALU stage, which is on the critical path.

This multiplexer can be moved into the memory stage. Note that on-chip memories
(scratchpad or cache) are faster than an ALU, including forwarding.
This leads to a 4-stage pipeline and removes the forwarding path from
WB to EXE. Thus reducing the critical path.

\subsection{Sharing an Adder for Address Calculation?}

In the classic RISC architecture, the address for a memory operation
is computed in the EX stage. The adder in the ALU is reused to add an
immediate value to a register value to compute the effective address.

This sharing of the adder was probably a good design tradeoff in the 80's
where resources were scarce.
However, today, adders are cheap (and fast). A 32-bit adder uses 32 LCs in an FPGA.
This is in the range of about 2\,\% of a complete pipeline implementation.
Since addition operations are inexpensive and fast, there is no necessity to 
reuse the ALU for address calculation.
Therefore, we propose to have a dedicated adder for the address calculation.

As reading from the register file is fast, we can compute the memory address in the decode stage.
Then, we can merge the execution stage with the memory stage.
This configuration results in a 3-stage pipeline and eliminates another forwarding path.
Another benefit of having memory access parallel to the ALU at the same stage
is the avoidance of load-use hazards. Data loaded from memory will be available
in the next clock cycle and forwarded to the ALU input.

Decoding instructions, particularly those in the RISC architecture, require a
relatively small number of gates, and decoding is faster than an ALU operation. 
Thus, it becomes a viable consideration to relocate that multiplexer
responsible for selecting between the register file value and the immediate value
into the decode stage. This proposed adjustment represents an additional measure
to minimize the critical path within the execution stage.

\begin{figure*}[t]
  \centering
  \includegraphics[width=\textwidth]{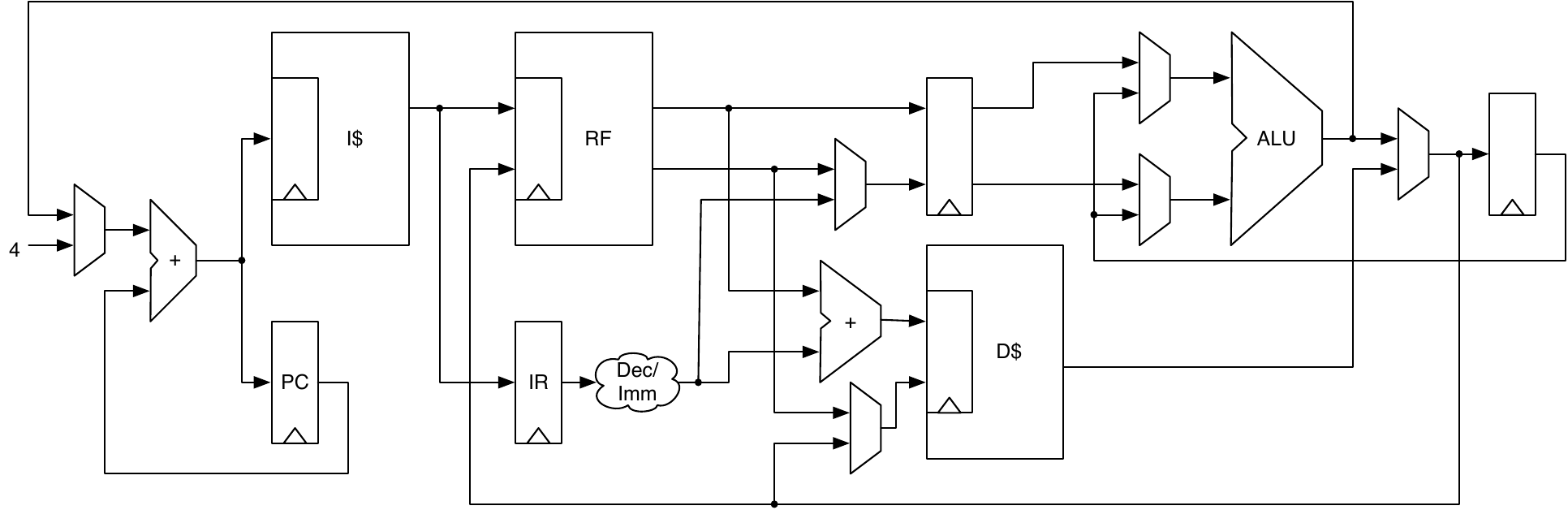}
  \caption{A 3-stage RISC-V processor pipeline.}
  \label{fig:3-stages}
\end{figure*}

Figure~\ref{fig:3-stages} shows a simplified schematic of the resulting 3-stage RISC-V pipeline.
Memory access is merged with a possible ALU operation in the EX stage. One forwarding path
from the ALU result or memory read to the next ALU operation exists.
Another forwarding path is from the combinational ALU result or memory read to the memory
input.

\subsection{Branching and Load-Use Hazard}

When we assume branch decisions and the branch destination computation
in EX, a taken branch takes three clock cycles. The branch can be reduced to two clock cycles
by moving the branch decision and computation into the ID stage or implementing a branch predictor.
This exploration is out of the scope of this paper.

The 4- and 5-stage pipelines suffer from the so-called load-use hazard.
A load-use hazard occurs when an instruction following a load instruction depends
on the result of the load. This result cannot be forwarded. We need to stall the pipeline by one clock
cycle (or restart the pipeline with the use instruction).
Note that the 3-stage pipeline does not suffer from that load-use hazard.

\subsection{Is the Proposed Pipeline Organization Controversial?}

We are aware that the classic RISC pipeline is considered to have
five stages. We explored several textbooks and slides from computer architecture
courses, and they have one thing in common: the RISC pipeline shown consists of
five stages.

I have taught computer architecture for around 15 years and have always
shown that classic pipeline organization. I even joked with the students that when
I wake them up in the middle of the night; they should be able to name those five
stages.

Maybe this paper is controversial. Maybe only the teaching is stuck
in the ``one and only'' classic pipeline organization? In related work, we will see that other organizations exist in the wild.

It might be controversial to question about 40-year-old common wisdom.
However, this paper tries to do so.

This paper argues that the 5-stages are probably not the sweet spot
for short-pipeline organizations. This paper does not argue against longer pipelines.
It argues for shorter and easier pipelines as a starting point.
There will be benefits of 6, 7, or 8 stages in-order pipelines.
We consider exploring those benefits and comparing our design with longer pipelines as future work.

\section{Implementation}
\label{sec:impl}

This project aims to provide examples of RISC pipeline implementations for education
in computer architecture. Therefore, the code focuses on readability
and avoiding distractions by performance or size optimizations.
However, we think clean code does not need to result in a large design with
a lower clock frequency.

For the design, we use Chisel~\cite{chisel:dac2012} \cite{chisel:book}, a hardware construction
language embedded in Scala. Chisel is a bit more high-level than VHDL or Verilog.
However, the more important fact is that it results in better readable code.
Furthermore, we switched our education in digital design to Chisel in 2020.
Our students are used to writing and reading Chisel code.

\subsection{A Simulator}

Each non-trivial digital design project should start with a simulation of that design.
A processor is a good example. In our computer architecture course, students must write 
as a final project a RISC-V ISA simulator.
Writing a simulator is a good preparation for an optional course of
implementing a RISC-V processor in an FPGA.

We wrote a simple instruction set simulator of RV32I. 
As the pipeline design was planned to be written in Chisel, we used Scala for the simulator.
This has two benefits: (1) we can co-simulate the hardware design and
the ISA simulator in the same Java virtual machine; and (2) we can share
constants between the simulator and the hardware.
The ISA simulator is around 300 lines of readable code.

\subsection{The Baseline}

As a baseline, we implemented the proposed 3-stage pipeline as described before.
We aimed for reusable code and put common functionality into Chisel functions, such as decoding,
immediate generation, and the ALU code.
In Chisel, hardware generators, including circuits that contain registers, can be put into plain functions.

The current implementation follows the description of the proposed 3-stages
pipeline from the former section. Memory addresses are computed in ID.
Branch decision and destination address generation are located in EX.
Immediate constant generation is performed in ID.
However, the selection between the register value and the immediate is placed in EX.
In future work, we will explore variations in the placement of parts between ID and EX.

The code for the 3-stages pipeline is around 140 lines of code, and additional
250 lines of code for decode, immediate generation, and the ALU are placed in functions to be reused by the other implementations.
The number of lines of code is not the only measurement for readability; however, browsing around 400 lines of code is more manageable than reading 3000 lines of code.

We will let others (our students) comment on the readability.
Future work will also be to work on the readability and clearness of the code.

\subsection{Extending the Baseline}

Extending the 3-stage pipeline towards 4 and 5 stages was straightforward.
However, this resulted in quite a lot of copied code, which is not
good.\footnote{Doug Locke stated: ``Two copies are never the same''.}
Our future work will explore ways to share more code without distracting
readability.

%

\section{Evaluation}
\label{sec:eval}

We evaluate the three different designs in two FPGA families and the OpenLane2
open-source ASIC flow for the open-source PDK from SkyWater in 130\,nm technology.
For the evaluations with FPGAs, we have set the timing constraint for the clock
to 200\,MHz to push the synthesis tools to optimize for speed.
The ASIC flow with OpenLane will not finish when the timing constraint is not met. Therefore, we set it to a frequency of 50\,MHz.

After placing and routing in the relevant tools, we derive the results from the reports.
We can find those results quickly in the reports using the Quartus (Altera) and Vivado (AMD) tools. The OpenLane flow is a bit more elaborated, consisting of 72 steps
with a report and log file for every step.
We find the size of the design in report 12 (\code{floorplan}) when the floor plan is generated.
We can find the slack relative to the requested frequency in report 51 (\code{stapostpnr}).
We report the computed maximum frequency for the chip at $25^\circ$C
at 1.8~V.
For the Altera FPGA, we report the timing results for the slow model at $85^\circ$C
and 1.8~V.
Vivado offers only one timing result.

To provide a baseline, we synthesized two designs targetting Cyclone~IV: (1) on-chip memory
with registers on the input and output, including forwarding for reads and writes
to the same address and (2) our ALU of the 3-stage design, including all forwarding paths and multiplexers for immediate versus register value, the PC for
JAL(R), and some more multiplexers.
An on-chip memory of 4~KB can be clocked at 587.89~MHz and a 16~KB version
at 272.33~MHz. In both cases, Quartus limits the report to 250~MHz due to restrictions
of the maximum I/O toggle rate to 250~MHz.
The ALU, with inputs and outputs connected to registers, is reported to have a maximum
clock frequency of 96.05~MHz.
This shows that on-chip memory (in an FPGA) is faster than ALU operations.
This observation motivated our pipeline to include some logic after the on-chip memory output, e.g., operations on the register file output and a 2:1 multiplexer at the data memory output.



As we are interested in the pipeline comparison alone, we leave out
any instruction and data memories or caches for the synthesis.
Our implementation contains two memory interfaces: (1) one for the instruction
memory and (2) one for the data memory.
A complete processor can attach caches, scratchpad memories, or a shared
memory interface.
For testing the processor (in simulation and in an FPGA), we use scratchpad memories.

The classic performance equation for in-order pipelines is the instruction count
multiplied by the clocks per instruction (CPI) multiplied by the clock period,
resulting in execution time.
The instruction count for all programs is the same as we use the same ISA.
The value of clocks per instruction (CPI) with real benchmarks is missing.
However, all three pipelines have a very similar CPI: instructions have a CPI of
one, a taken branch has two clock cycles penalty, and on the 4- and 5-stages
pipelines, there is one clock cycle penalty for a load-use hazard.
The main source of a higher CPI is missing in the caches, which we do not compare
in the paper.
Therefore, for the performance comparison, we use the maximum clock frequency
as a proxy for the performance.

\subsection{FPGA Results} 

For the evaluation, we used an Intel Cyclone IV FPGA. Although a bit dated, it is the
FPGA is on the popular Terrasic DE2-115 FPGA board we use at our university.
Furthermore, several designs are available with published numbers for this FPGA
for a rough comparison.
The Java processor JOP~\cite{jop:jnl:jsa2007} needs 2050 LCs for the implementation
and can be clocked at 100 MHz.
The time-predictable processor Patmos~\cite{patmos:rts2018} needs 7602~LCs
for the implementation and can be clocked at 81.7~MHz.
Patmos is also a RISC processor, and the critical path is in the ALU
including the forwarding multiplexers.

We synthesize two versions for each design and FPGA: one with the register file
built out of flip-flops (FF) and one using synchronous on-chip memory (mem).
Table~\ref{tab:altera} shows the results for the three pipeline organizations
for the Altera FPGA. The logic resource usage is reported in logic elements (LE),
which consist of a 4-bit lookup table and a flip-flop. The number of flip-flops is
also noted in the column Register. RAM bits are the utilization of the on-chip memories.
2048 bits of memory (two memory blocks) are needed to support two read ports for the register file.
We observe that using on-chip memory for the register file results 
in a faster implementation in most cases. And the resource usage of LE is reduced.
However, the most surprising observation is that most 4- and 5-stage pipelines
have a \emph{lower} maximum clocking frequency. This contradicts the very
purpose of pipelining, which is to increase the maximum clocking frequency.
Our explanation is that the ALU, including the forwarding multiplexers, is the critical
path in FPGA technology (memories are fast). Therefore, longer pipelines adding
forwarding multiplexers hurt the maximum clock frequency (up to 5 stages).

\begin{table}
\centering
\caption{Wildcat results for the Altera FPGA Cyclone~IV}
\begin{tabular}{lrrrr}
\toprule
Design (Cyclone IV) & max. frequency & logic elements& flip-flops & RAM bits \\
\midrule
 Three stages (FF)  & 80.2 MHz & 3,130 & 1,295 & 0        \\
 Three stages (mem) & 86.2 MHz & 1,756 & 379   & 2,048    \\
 Four stages (FF)   & 83.9 MHz & 3,040 & 1,367 & 0        \\
 Four stages (mem)  & 84.5 MHz & 1,727 & 451   & 2,048    \\
 Five stages (FF)   & 78.4 MHz & 3,107 & 1,438 & 0        \\
 Five stages (mem)  & 75.7 MHz & 1,813 & 522   & 2,048    \\
\bottomrule
\end{tabular}
\label{tab:altera}
\end{table}

Furthermore, we synthesize the designs for the AMD (former Xilinx) FPGA Artix~7,
the FPGA on the Nexys~A7 board. We used the Nexys board to run small
example programs on Wildcat on real hardware.
Table~\ref{tab:xilinx} shows the results for the Artix FPGA.
A logic cell (LC) in the Artix FPGA consists of a 6-bit lookup table and a register.
Therefore, the numbers are different from the Altera FPGA.
This FPGA is a newer generation than the Cyclone~IV FPGA, so we observe higher
possible clock frequencies. Again, we observe the general trend of slower clock
frequency with longer pipelines.
The memory-based register file is implemented as a distributed
LUT RAMs in the Artix FPGA. Therefore, the number of bits in the block
RAMs are zero. Of the 1256 LUTs in the three stages pipeline,
48 are used as distributed RAMs for the register file.

\begin{table}
\centering
\caption{Wildcat results for the AMD (former Xilinx) FPGA Artix~7}
\begin{tabular}{lrrrr}
\toprule
Design (Artix 7) &  max. frequency & logic cells & flip-flops & RAM bits \\
\midrule
 Three stages (FF)  & 99.6 MHz  & 1,744 & 1,329 & 0        \\
 Three stages (mem) & 112.3 MHz & 1,270 & 303   & 0        \\
 Four stages (FF)   & 107.5 MHz & 1,551 & 1,438 & 0        \\
 Four stages (mem)  & 111.2 MHz & 993   & 442   & 0        \\
 Five stages (FF)   & 106.1 MHz & 1,724 & 1,511 & 0        \\
 Five stages (mem)  & 102.0 MHz & 1,158 & 515   & 0        \\
\bottomrule
\end{tabular}
\label{tab:xilinx}
\end{table}

\subsection{ASIC Results}

We use the SkyWater\footnote{\url{https://www.skywatertechnology.com/}} project for the ASIC evaluation.
SkyWater is a fab with a 130\,nm process and together with efabless\footnote{\url{https://efabless.com/}}
they provide a service to run a multi-project wafer (MPW) design.
If the design is open source, Google sponsored the chip production, packaging,
and a PCB. However, this sponsorship has been stopped at the time of this writing.
Therefore, we plan to submit Wildcat to the next run of Tiny Tapeout~\cite{TinyTapeout}, a very affordable way to produce an ASIC.

Although the SkyWater130 process is an old technology, it serves well for comparisons.
The process development kit (PDK) of SkyWater130 is available in open source, providing
an equal playground for many projects. We expect that the SkyWater130 process will become a reference platform for future publications in computer architecture research.

As we do not have a memory compiler available for this process, we can only
implement the register file in flip-flops.
Table~\ref{tab:asic} shows the results for the SkyWater130 ASIC.
The ASIC flow results confirm the trend we see in the FPGA results.
As expected, the area increases with larger pipelines, and the maximum clock frequency
decreases, which was unexpected. 

To set the area of the pipeline into context, the area is about 0.2~mm2.
A multi-project waver tile with a user area of 10~mm2 is available for around \$\,10,000
within the efabless\footnote{\url{https://efabless.com/prototyping}} project.

\begin{table}
\centering
\caption{Wildcat results with the SkyWater130 ASIC process}
\begin{tabular}{lrrrr}
\toprule
Design (SkyWater130) & fmax (MHz) & Size \\
\midrule
Three stages (FF) & 81.2 MHz & 429 x 432 \text{\textmu}mm$^2$ \\
Four stages (FF)  & 73.2 MHz & 433 x 438 \text{\textmu}mm$^2$  \\
Five stages (FF)  & 69.5 MHz & 439 x 443 \text{\textmu}mm$^2$  \\
\bottomrule
\end{tabular}
\label{tab:asic}
\end{table}

\section{Related Work}
\label{sec:related}

David Patterson and John Hennessy coined the term \emph{reduced instruction set computer} (RISC).
The first publicly available RISC processor was the RISC I~\cite{risc:1980} \cite{risc1:patterson:1981},
designed at UCB and MIPS~\cite{MIPS:1982}, designed at Stanford University.
The IBM 801 predates the two RISC processors from academia, but details have not been disclosed
in the early 80's.


The RISC-V project started with Andrew's thesis~\cite{Waterman:EECS-2016-1}.
To verify his work on defining \emph{the} RISC instruction set, the research group in Berkeley
developed several versions of a RISC-V microprocessor. The initial version was an
instruction set simulator called Spike. The hardware implementation following the simulator
was called Rocket~\cite{rocket:techrep}. Rocket is implemented in Chisel and represents a
5-stage in-order scalar pipeline. The project is also called the Rocket Chip generator, as it is a collection
of tools to generate RISC-V-based system-on-chips. Besides
the Rocket in-order pipelined core, the generator contains components
with the TileLink protocol, several tools, and Diplomacy~\cite{cook2017diplomatic}.
Diplomacy is an extension package for Chisel, providing support for two-phase
hardware elaboration. This approach enables dynamic negotiation of specific
parameters between modules.
Rocket was not intended as a reference implementation of a RISC-V microprocessor,
but is considered as a reference by many.

As Rocket was a bit too advanced for education, a group in Berkeley developed the
Sodor family of RISC-V processors.
The Sodor project\footnote{\url{https://github.com/ucb-bar/riscv-sodor}} is an open-source initiative that contains educational RISC-V processor cores.
The cores range from a single-stage, 2-stage, 3-stage, up to a classic 5-stage version.
Sodor is written in Chisel. However, some of the code is quite advanced, such as Scala code,
which is not an easy read for a beginner of computer architecture and Chisel.
Furthermore, the Sodor project is no longer self-contained. It needs the
The Chipyard SoC generator itself has an elaborate setup.
In contrast to Sodor, Wildcat is available in a standalone, and we have put effort
into producing readable code instead of showing off with clever but hard-to-read solutions.
Furthermore, Sodor cannot easily be used in an FPGA, as it depends on asynchronous memories:

\begin{quotation}
All processors talk to a simple scratchpad memory (asynchronous, single-cycle), with no backing outer memory.... Programs are loaded in via a Host-target Interface (HTIF) port (while the core is kept in reset), effectively making the scratchpads 3-port memories (instruction, data, HTIF).
\end{quotation}

YARVI (Yet Another RISC-V Implementation)~\footnote{\url{https://github.com/tommythorn/yarvi}} by Tommy Thorn was probably the first RISC-V implementation that could be synthesized into an FPGA (originally released in 2014).
The first implementation was a multi-cycle version followed by a pipelined version.
The current version of YARI, called YARVI2,  is an 8-stage pipeline with an effort on branch prediction.
The website reports a maximum clock frequency of over 100 MHz on a Cyclone~V.
YARVI, with its 8-stage in-order pipeline, shows a path for future versions of Wildcat.

PULPino~\cite{PULPino} is a 32-bit RISC-V microcontroller system developed at ETH Zurich.
It is written in SystemVerilog in a conservative style (e.g., not using structures but
individual signals.) We appreciate that project but consider it too complex for education or research.

Ibex~\cite{Ibex2017}\footnote{\url{https://github.com/lowRISC/ibex}} is a two-stage pipeline
with an additional clock cycle for memory access. Therefore, similar in design to our
Wildcat project. The pipeline can be extended with a write-back stage. The documentation
states:

\begin{quotation}
Ibex can be configured to have a third pipeline stage (Writeback) which has major effects on performance and instruction behavior. The details of its impact are not yet documented here.
\end{quotation}

PicoRV32~\footnote{\url{https://github.com/YosysHQ/picorv32}} is a RISC-V implementation optimized for small size and a high clocking frequency but not for execution speed. The application area is as an auxiliary processor in FPGA or ASIC designs. The focus on high clock frequency shall simplify the integration into existing designs without the need for clock domain crossing. The implementation is sequential, with instructions taking between 3 and 14 clock cycles.
The single Verilog file picorv32.v is about 3000 lines of Verilog code, which is not an easy read.

RISC-V processors are specialized for different domains. For example,
MINOTAuR~\cite{gruin2021speculative} is a time-predictable RISC-V core aiming to be used
in real-time systems. Although Wildcat's focus is currently as an educational RISC-V core,
we plan to add features from the T-CREST platform~\cite{t-crest:dasia:2014} to use it in real-time systems.

Ripes~\cite{ripes} is a graphical simulator for different configurations of a RISC-V pipeline.
Morten developed Ripes while he was still a student at DTU. We use Ripes to educate students in computer architecture.
Ripes seems to have become the most used RISC-V simulator to teach computer architecture.
Another RISC-V simulator is WebRISC-V~\cite{giorgi2019webriscv}. The simulator was originally
developed for MIPS and has been adapted to the RISC-V instruction set.

There are many RISC-V implementations, both open-source and commercial closed-source. Describing them all would result in a long survey paper.
A small set of application class RISC-V implementations has been compared~\cite{dorflinger2021comparative} concerning performance, area, and power.

\subsection{Source Access}

Wildcat is available in open source on GitHub:
\url{https://github.com/schoeberl/wildcat}.

\section{Conclusion}
\label{sec:conclusion}

This paper presents Wildcat, a RISC-V implementation aiming for simplicity and to be
used in education. We designed and coded three RISC pipelines with 3, 4, and 5 stages.
The 3-stage pipeline is superior to the other version not only because of lower
resource requirements but also because of higher performance (maximum frequency).
This result is counterintuitive, as pipelining aims for higher frequencies.
However, longer pipelines need more forwarding paths to the ALU, which hurts the critical path
in the execution stage.
We conclude that the classic 5-stages pipeline organization, taught in
computer architecture classes worldwide, is not the sweet spot for short
pipeline organizations. Neither for teaching nor for small embedded processors.

\subsection*{Acknowledgment}

I would like to thank Tommy Thorn for the ongoing, inspiring, and enjoyable discussions
of RISC pipeline organizations.

\bibliographystyle{splncs04}
\bibliography{msbib, arcs}

\begin{thebibliography}{10}
\providecommand{\url}[1]{\texttt{#1}}
\providecommand{\urlprefix}{URL }
\providecommand{\doi}[1]{https://doi.org/#1}

\bibitem{rocket:techrep}
Asanovic, K., Avizienis, R., Bachrach, J., Beamer, S., Biancolin, D., Celio,
  C., Cook, H., Dabbelt, D., Hauser, J., Izraelevitz, A., Karandikar, S.,
  Keller, B., Kim, D., Koenig, J., Lee, Y., Love, E., Maas, M., Magyar, A.,
  Mao, H., Moreto, M., Ou, A., Patterson, D.A., Richards, B., Schmidt, C.,
  Twigg, S., Vo, H., Waterman, A.: The rocket chip generator. Tech. Rep.
  UCB/EECS-2016-17, EECS Department, University of California, Berkeley (Apr
  2016)

\bibitem{chisel:dac2012}
Bachrach, J., Vo, H., Richards, B., Lee, Y., Waterman, A., Avizienis, R.,
  Wawrzynek, J., Asanovic, K.: Chisel: constructing hardware in a scala
  embedded language. In: The 49th Annual Design Automation Conference (DAC
  2012). pp. 1216--1225. ACM, San Francisco, {CA}, {USA} (June 2012)

\bibitem{cook2017diplomatic}
Cook, H., Terpstra, W., Lee, Y.: {Diplomatic Design Patterns: A TileLink Case
  Study}. In: 1st Workshop on Computer Architecture Research with RISC-V.
  vol.~23 (2017)

\bibitem{Ibex2017}
Davide~Schiavone, P., Conti, F., Rossi, D., Gautschi, M., Pullini, A., Flamand,
  E., Benini, L.: Slow and steady wins the race? a comparison of
  ultra-low-power risc-v cores for internet-of-things applications. In: 2017
  27th International Symposium on Power and Timing Modeling, Optimization and
  Simulation (PATMOS). pp.~1--8. \doi{10.1109/PATMOS.2017.8106976}

\bibitem{dorflinger2021comparative}
D{\"o}rflinger, A., Albers, M., Kleinbeck, B., Guan, Y., Michalik, H., Klink,
  R., Blochwitz, C., Nechi, A., Berekovic, M.: A comparative survey of
  open-source application-class risc-v processor implementations. In:
  Proceedings of the 18th ACM International Conference on Computing Frontiers
  (2021)

\bibitem{PULPino}
Gautschi, M., Schiavone, P.D., Traber, A., Loi, I., Pullini, A., Rossi, D.,
  Flamand, E., Gürkaynak, F.K., Benini, L.: Near-threshold risc-v core with
  dsp extensions for scalable iot endpoint devices. IEEE Transactions on Very
  Large Scale Integration (VLSI) Systems  \textbf{25}(10),  2700--2713 (2017).
  \doi{10.1109/TVLSI.2017.2654506}

\bibitem{giorgi2019webriscv}
Giorgi, R., Mariotti, G.: Webrisc-v: A web-based education-oriented risc-v
  pipeline simulation environment. In: ACM Workshop on Computer Architecture
  Education (WCAE-19) (2019)

\bibitem{gruin2021speculative}
Gruin, A., Carle, T., Rochange, C., Cass{\'e}, H.: Speculative execution and
  timing predictability in an open source risc-v core. In: IEEE Real-Time
  Systems Symposium (RTSS) (2021)

\bibitem{harris2021digital}
Harris, S., Harris, D.: Digital Design and Computer Architecture, RISC-V
  Edition. Elsevier Science (2021)

\bibitem{Hennessy06}
Hennessy, J., Patterson, D.: Computer Architecture: {A} Quantitative Approach,
  4th ed. Morgan Kaufmann Publishers (2006)

\bibitem{MIPS:1982}
Hennessy, J.L., Jouppi, N.P., Baskett, F., Gill, J., Towle, R.: Mips: A
  microprocessor architecture pp. 17--22 (1982). \doi{10.1145/1014194.800930}

\bibitem{risc:1980}
Patterson, D.A., Ditzel, D.R.: The case for the reduced instruction set
  computer

\bibitem{risc1:patterson:1981}
Patterson, D.A., Sequin, C.H.: {RISC I}: A reduced instruction set {VLSI}
  computer. In: Proceedings of the 8th annual symposium on Computer
  Architecture. pp. 443--457. ISCA '81, IEEE Computer Society Press, Los
  Alamitos, CA, USA (1981)

\bibitem{ripes}
Petersen, M.B.: Ripes: A visual computer architecture simulator. In: 2021
  ACM/IEEE Workshop on Computer Architecture Education (WCAE) (2021)

\bibitem{jop:jnl:jsa2007}
Schoeberl, M.: A {Java} processor architecture for embedded real-time systems.
  Journal of Systems Architecture  \textbf{54/1--2},  265--286 (2008).
  \doi{http://dx.doi.org/10.1016/j.sysarc.2007.06.001}

\bibitem{chisel:book}
Schoeberl, M.: Digital Design with Chisel. Kindle Direct Publishing (2019),
  available at \url{https://github.com/schoeberl/chisel-book}

\bibitem{wildcat:2024}
Schoeberl, M.: The educational risc-v microprocessor wildcat. In: Proceedings
  of the Sixth Workshop on Open-Source EDA Technology (WOSET) (2024)

\bibitem{patmos:rts2018}
Schoeberl, M., Puffitsch, W., Hepp, S., Huber, B., Prokesch, D.: Patmos: A
  time-predictable microprocessor. Real-Time Systems  \textbf{54(2)},  389--423
  (April 2018). \doi{10.1007/s11241-018-9300-4}

\bibitem{t-crest:dasia:2014}
Schoeberl, M., Silva, C., Rocha, A.: {T-CREST}: A time-predictable multi-core
  platform for aerospace applications. In: Proceedings of Data Systems In
  Aerospace (DASIA 2014). Warsaw, Poland (June 2014)

\bibitem{TinyTapeout}
Venn, M.: Tiny tapeout: A shared silicon tape out platform accessible to
  everyone. IEEE Solid-State Circuits Magazine  \textbf{16}(2),  20--29 (2024).
  \doi{10.1109/MSSC.2024.3381097}

\bibitem{Waterman:EECS-2016-1}
Waterman, A.: Design of the RISC-V Instruction Set Architecture. Ph.D. thesis,
  EECS Department, University of California, Berkeley (Jan 2016)

\end{thebibliography}

\end{document}